\documentclass[twocolumn,pra,showpacs,unsortedaddress]{revtex4-1}

\usepackage{graphicx, color}
\usepackage{mathrsfs}
\usepackage{amsmath, amssymb}

\newcommand{\ket}[1]{\ensuremath{\lvert #1 \rangle}}
\newcommand{\bra}[1]{\ensuremath{\langle #1 \rvert}}
\newcommand{\bracket}[2]{\ensuremath{\langle #1 \vert #2 \rangle}}

\newcommand{\ud}{\ensuremath{\:\mathrm{d}}}
\newcommand{\tqa}{\ensuremath{\tau_{\mathrm{QSL}}}}
\newcommand{\tqac}{\ensuremath{T_{\mathrm{QSL}}}}
\newcommand{\tqn}{\ensuremath{T_{\mathrm{QSL}}^\ast}}

\begin{document}

\title{Communication at the quantum speed limit along a spin chain}

\author{Michael~Murphy}
\email{michael.murphy@uni-ulm.de}
\author{Simone~Montangero}
\affiliation{Institut f\"{u}r Quanteninformationsverarbeitung, Universit\"{a}t Ulm, Albert-Einstein-Allee 11, D-89069 Ulm, Germany}

\author{Vittorio~Giovannetti} 
\affiliation{NEST-CNR-INFM \& Scuola Normale Superiore, P.zza dei Cavalieri 7,56126 Pisa, Italy}

\author{Tommaso~Calarco}
\affiliation{Institut f\"{u}r Quanteninformationsverarbeitung, Universit\"{a}t Ulm, Albert-Einstein-Allee 11, D-89069 Ulm, Germany}


\begin{abstract}

  Spin chains have long been considered as candidates for quantum
  channels to facilitate quantum communication. We consider the transfer
  of a single excitation along a spin-1/2 chain governed by
  Heisenberg-type interactions. We build on the work of Balachandran and
  Gong \cite{Balachandran2008}, and show that by applying optimal
  control to an external parabolic magnetic field, one can drastically
  increase the propagation rate by two orders of magnitude. In
  particular, we show that the theoretical maximum propagation rate can
  be reached, where the propagation of the excitation takes the form of
  a dispersed wave. We conclude that optimal control is not only a
  useful tool for experimental application, but also for theoretical
  enquiry into the physical limits and dynamics of many-body quantum
  systems.

\end{abstract}


\maketitle


\section{Introduction}


Quantum computers promise to allow efficient simulation of large dynamic
and complex systems and deliver performance advantages over their
classical counterparts. One of the central considerations for the
construction of a quantum computer is an infrastructure that can rapidly
and robustly transport qubit states between sites where qubit operations
can be performed.  The components for this infrastructure may be though
of as quantum channels for quantum information transfer. One of the
technologies under investigation to constitute such a channel is the
one-dimensional spin-chain \cite{Bose2007, Balachandran2008,
  Romero-isart2007, Christandl2004, Subrahmanyam2004, Bose2003, Kay2006,
  Burgarth2009, Burgarth2007, Giovannetti2006}, which consists of a
string of particles coupled via their spin degrees of freedom, each
acting as an effective two-level quantum system. As is customary in
quantum information processing, proper engineering of the control
parameters of the system is essential to achieve the high fidelity
necessary for robust quantum computation. This can be obtained, for
instance, by employing a numerical optimisation method which, for the
specific settings of the problem, seeks the optimal control pulses that
allow one to implement the desired operation
\cite{Calarco2004,Montangero2007,DeChiara2008,Schulte-Herbruggen2005,Khaneja2005,Grace2007,Nebendahl2009,Sporl2007,Schirmer2009a,Nigmatullin2009,Tesch2002}.
%
%
In this paper, we apply such a method, known in the literature as the
Krotov method
\cite{Sklarz2002,Tannor1992,Somloi1993,Palao2002,Palao2003}, to the
case of quantum state transfer along a one-dimensional spin chain. The
specific system we use was introduced by Balachandran and Gong
\cite{Balachandran2008}, but here we show that by designing the external
driving parameters with optimal control methods, one can obtain a
significant increase in fidelity, even over short time scales
\cite{Schirmer2009}.

These high-fidelity, high-speed transmissions exhibit interesting
characteristics. If one ignores the effects taking place near the
boundaries, the evolution of the excitation is that of a dispersed wave,
moving with almost constant velocity along the chain. This velocity is
independent of the chain length, and furthermore has an upper bound,
indicating the presence of a fundamental limit on the rate of
transmission. Through a closer analysis, we show that this limit can be
directly related to the theoretical maximum speed of the state transfer
allowable by the laws of quantum mechanics \cite{Yung2006,
  Giovannetti2003, Giovannetti2004, Pfeifer1993, Lloyd2000,
  Caneva2009}.

Producing time-optimal gates has already been explored in the literature
\cite{Carlini2006, Carlini2007, Khaneja2001, Khaneja2002, Reiss2003,
  Fisher2009} where the authors considered geodesics on the Bloch sphere
for systems with a low number of dimensions. Unfortunately, extension of
these methods to many-body systems (such as the system we consider here)
are prohibitively difficult. Conversely, the numerical optimisation
methods that we employ have little difficulty in finding sets of optimal
solutions, even at this limit. In effect, we demonstrate that through
application of optimal control, we can not only transmit the excitation
with a high fidelity, but also at the fastest possible speed. One can
even reverse the problem, implying that optimal control can be used to
probe such fundamental dynamical limits on many-body quantum
systems. Such tools will be invaluable as the ambition of quantum
science leads it towards investigations of systems of greater
complexity which are less tractable analytically.

The paper is arranged as follows. In Section~\ref{sec:sc}, we describe
the system used for information transfer in more detail and the precise
scheme which we will use for propagating quantum information in the
system. Section~\ref{sec:oct} discusses the application of optimal
control to the transfer scheme, and shows that optimal control can
effect significant gains in the transfer speed. We then discuss the
fundamental limit of these improvements in Section~\ref{sec:conn}, and
show that optimal control in fact allows us to reach this limit, thus
allowing us to transfer the spin state in the fastest possible time
allowable by the laws of quantum mechanics. Finally, we present the
conclusions in Section~\ref{sec:conc}.


\section{Spin chains as quantum channels}\label{sec:sc}

\subsection{Overview}
Using spin chains as quantum channels for communication between two
parties was first proposed by Bose in 2003 \cite{Bose2003} and later
developed in a series of papers (we refer the reader to
Ref.~\cite{Bose2007} for a review). The idea is relatively simple: Alice
(the sender) has a quantum state she wants to relay to Bob (the
receiver). Between them is a one-dimensional chain of $N$ spin-1/2
particles which are coupled via nearest-neighbour interactions. Alice
has access to the first spin in this chain, and can prepare its spin
state as she chooses. Bob has access to the final site (the terms `spin'
and `site' will be used interchangeably), whose state he can read
out. Following \cite{Balachandran2008}, we apply an external parabolic
magnetic field, which Alice can control. The procedure for sending
quantum information along the chain is as follows.
\begin{enumerate}
\item The spin chain is prepared in its ground state with respect to
  the external magnetic field.
\item Alice prepares the initial spin state to be the state she wishes
  to transfer.
\item By manipulating the magnetic field, Alice controls the propagation
  of the spin along the chain, which takes place due to the coupling
  between the spin degrees of freedom.
\item After some prescribed time when the state has been transferred
  to the final site, Bob reads out the state of this site.
\end{enumerate}

\subsection{The Hilbert space and Hamiltonian}
\begin{figure}[tb]
  \begin{center}
    \includegraphics{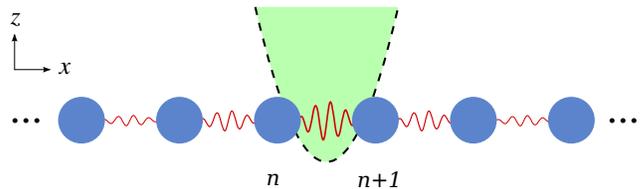}
  \end{center}
  \caption{(Colour online) The one-dimensional spin chain used for
    information transfer. The (blue) filled circles represent sites
    along the chain, with the applied magnetic field depicted. The
    effective couplings are indicated operating between the sites.}
  \label{fig:sc1}
\end{figure}

The model we consider is sketched in Fig.~\ref{fig:sc1}. It is composed
of a one-dimensional spin-1/2 chain with $N$ sites, where distances are
measured by the variable $x$ (although this may not be a physical
distance). We will consider uniform Heisenberg nearest-neighbour
couplings characterised by the same coupling strength $J$, and the
presence of a parabolic external magnetic field in the $z$-direction,
normal to the direction $x$.  Consequently, the field will act on the
$n$th site as
\begin{equation}
  B_n(t) = C(t)\bigl(x_n - d(t)\bigr)^2\:,
\end{equation}
where $d(t)$ is the position of the field minimum along $x$ at time
$t$, and $C(t)$ is a measure of the global field strength.

The Hamiltonian then takes the form
\begin{equation}\label{eq:ham}
  H(t) = -\frac{J}{2}\sum_{n=1}^{N-1}\vec{\sigma}_n\cdot\vec{\sigma}_{n+1} + 
  \sum_{n=1}^{N}B_n(t)\sigma_n^z\:,
\end{equation}
where $n$ labels the spin sites, with $n = 1$ and $n = N$ referring to
the first and last spins, respectively, and $\vec{\sigma}_n =
(\sigma^x_n, \sigma^y_n, \sigma^z_n)$ are the Pauli spin operators for
the $n$th spin. For convenience, all system parameters are scaled to
make them dimensionless, and the coupling strength is set to $J =
1$. 

The dynamics are governed by the interplay between the nearest-neighbour
interactions and the interaction of each site with the external
parabolic magnetic field. When sites are far from the field minimum, the
local field strength dominates over the nearest-neighbour interactions,
effectively `switching off' the coupling between sites. For sites near
the minimum where the field is weak, the nearest-neighbour coupling
dominates, and the neighbouring sites interact with each other. These
two processes control the propagation of spin states along the chain.

\subsection{Communicating quantum information}
\begin{figure}[tb]
  \begin{center}
    \includegraphics{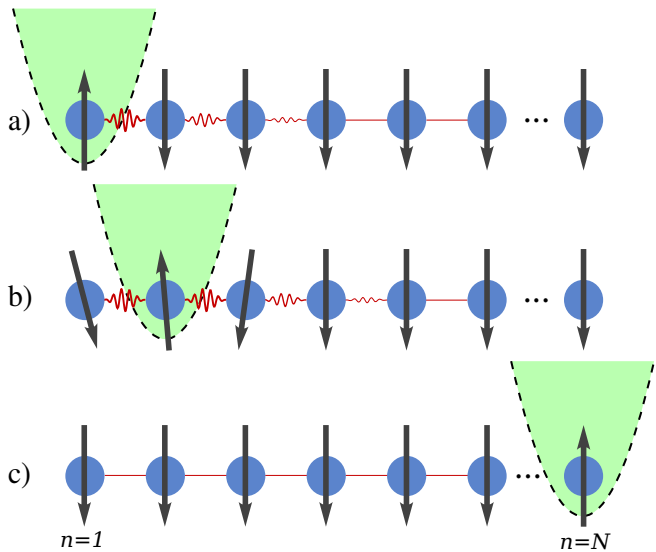}
  \end{center}
  \caption{(Colour online) The transfer begins with the state
    \ket{\Psi(0)} and with the potential minimum centred at $x = 0$. (a)
    The excitation is localised at the first site. (b) The field minimum
    moves along the chain during the evolution. (c) The spin state has
    been completely transferred to the final site in the chain.}
  \label{fig:sc2}
\end{figure}

We identify the computational basis for our system with the quantised
states of each spin, such that $\ket{0} = \ket{\downarrow}$ (spin down
with respect to $z$) and $\ket{1} = \ket{\uparrow}$ (spin up). Assume
that Alice prepares the chain in the initial state \ket{\Psi(0)}, with
the first spin site in the state \ket{1}, and all other sites to in
their ground state \ket{0}. We can write this state as
\begin{equation}
  \label{eq:psi0}
  \ket{\Psi(t = 0)} = \ket{\varphi_1} \equiv \ket{1} \otimes \ket{0} \otimes \ket{0} \otimes
  \dots \otimes \ket{0}\:,
\end{equation}
with the first spin site in the state \ket{1}, and all other sites to in
their ground state \ket{0}. The states $\ket{\varphi_n}$ are defined as
\begin{equation}
  \label{eq:phin}
  \ket{\varphi_n} \equiv \bigotimes_{m=1}^{N}
  \ket{\delta_{mn}}\:, \quad n = 1, \dots, N,
\end{equation}
where $\delta_{mn}$ is the Kronecker delta. Alice's goal is to
manipulate the magnetic field parameters $C(t)$ and $d(t)$ such that at
the final time $T$ the final state obeys 
\begin{equation}
  \label{eq:final_cond}
  |\bracket{\Psi(T)}{\varphi_N}|^2 = 1\:.
\end{equation}
The protocol for transferring the state is based on that described in
Ref.~\cite{Balachandran2008}, which we outline in
Figure~\ref{fig:sc2}. The transfer begins with the state \ket{\Psi(0)}
and with the potential minimum centred at $x = 0$. At first, the
interaction between the first two sites dominates over the interaction
with the locally weak magnetic field, and so the sites interact and the
spin state migrates from the first site to the second. As the field
minimum moves along the $x$-axis, nearest-neighbour interactions are
effectively switched on for pairs of spins closest to the minimum, and
switched off for spin pairs that are distant. By correctly moving the
field minimum and adjusting the field strength, the spin state is able
to traverse the chain.
The condition in Eq.~\eqref{eq:final_cond} means that we do not preserve
the phase of the initial state; to achieve this (as also discussed in
Ref.\cite{Balachandran2008}) one can use dual-rail encoding
\cite{Burgarth2005, Burgarth2005a, Burgarth2005b}, whereby one encodes
the qubit in the entanglement phase of a pair of spin chains. In what
follows, we shall only consider the phase-insensitive transfer of a
single excitation.

Since each site of the chain has two internal spin states, the size of
the Hilbert space $\mathcal{H}$ scales exponentially with the number of
sites, so that for $N (\geq 1)$ sites, $\mathrm{dim}\:\mathcal{H} =
2^N$. However, since $[H(t),\sum_{n=1}^{N} \sigma^z_n] = 0$, the state
of the system only evolves within the subspace $\mathcal{U} \subset
\mathcal{H}$, spanned by the $N$ basis states $\ket{\phi_n}$
\cite{Bose2007}. The reduced size of the effective Hilbert space is
particularly beneficial when one wants to numerically simulate the
evolution efficiently. We do this by solving the associated
Schr\"{o}dinger equation
\begin{equation}\label{eq:schrod}
  i \frac{\partial}{\partial t}\ket{\Psi(t)} = \hat{H}(t) \ket{\Psi(t)}\:,
\end{equation}
where $\hat{H}$ is the matrix form of the Hamiltonian that acts only on
the subspace $\mathcal{U}$
\begin{equation}\label{eq:ham2}
  \hat{H}(t) \equiv \hat{H}_0 + \hat{H}_1(t)\:, 
\end{equation}
with
\begin{equation}
  \hat{H}_0 =-2J + J\begin{pmatrix}
    1&1&0&\cdots&0&0\\
    1&0&1&&0&0\\ \vspace{-0.05in}
    0&1&0&&0&0\\ 
    \vdots&&&\ddots&&\vdots\\
    0&0&0&&0&1\\
    0&0&0&\cdots&1&1
  \end{pmatrix}
\end{equation}
and
\begin{equation}
\hat{H}_1(t) = \mathrm{diag}\left(f_0(t),f_1(t),\dots,f_{N-1}(t)\right)\:,
\end{equation}
where $f_n(t) = C(t)\left(x_n - d(t)\right)$ (note that we have rescaled
the energy so that spins pointing down do not contribute to the total
energy). The Schr\"{o}dinger equation is integrated numerically using
the Crank-Nicolson method \cite{Crank1996}.

This scheme was first considered by Balachandran and Gong
\cite{Balachandran2008}, who showed that by choosing $d(t) = st$ and
$C(t) = k$, where $s$ and $k$ are constant, one is able to adiabatically
transfer the state across the entire chain with relatively good
fidelities. However, the transfer rates here are very slow, with
transfer times typically on the order of $10^4J$ for fidelities greater
than 99\%.

In many proposed implementations of quantum computers, it is likely that
transport processes will take up a significant amount of the total
operating time. It therefore seems clear that one should seek to
minimise the time required for these processes. However, according to
quantum mechanics there is some fundamental limit which restricts the
speed at which we can communicate with our spin chain, referred to in
the literature as the `quantum speed limit' (QSL) \cite{Pfeifer1993,
  Lloyd2000, Giovannetti2003, Yung2006, Levitin2009, Caneva2009}. The
goal is to come as close as possible to this limit, effectively
communicating at the highest possible speed allowable by quantum
mechanics. We shall see in the next section that optimal control can
help us in this endeavour.


\section{Optimal dynamics}\label{sec:oct}

We can state our problem in the following way: we start with an initial
state, and want to control the system to produce the desired final
state. In our case, the initial state is $\ket{\varphi_1}$, and we want
to achieve the final state $\ket{\varphi_N}$ (up to a global phase). We
can control the evolution of the system using the external magnetic
field, in particular the time-dependent controls $d(t)$ and
$C(t)$. (Although in principle we could also control the inter-spin
coupling $J$, this is much more difficult to achieve experimentally.)
Optimal control theory provides us with a set of tools to search for the
optimal way to control the system, often referred to as the set of
optimal \emph{controls}.

\begin{figure}[tp]
  \begin{center}
    \includegraphics{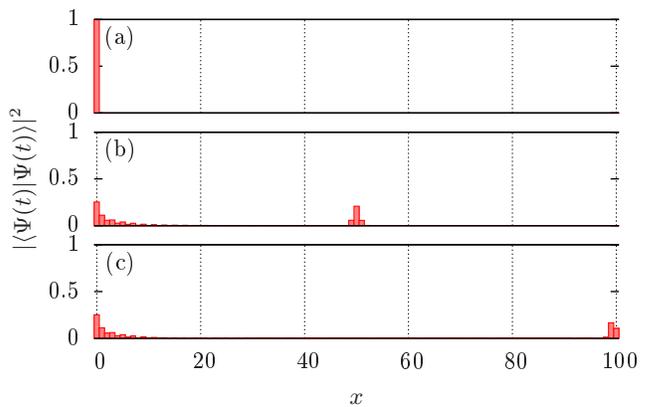}
  \end{center}
  \caption{(Color online) Excitation probability plotted against $x$ for
    a spin chain with $N = 101$ sites, at times (a) 0, (b) 100 and (c)
    200, in units of $J^{-1}$. Here, $d(t) = 0.5 t$ and $C(t) = 1$. The
    final fidelity is only around 15\%.}
  \label{fig:nocontpop}
\end{figure}

\begin{figure}[tp]
  \begin{center}
    \includegraphics{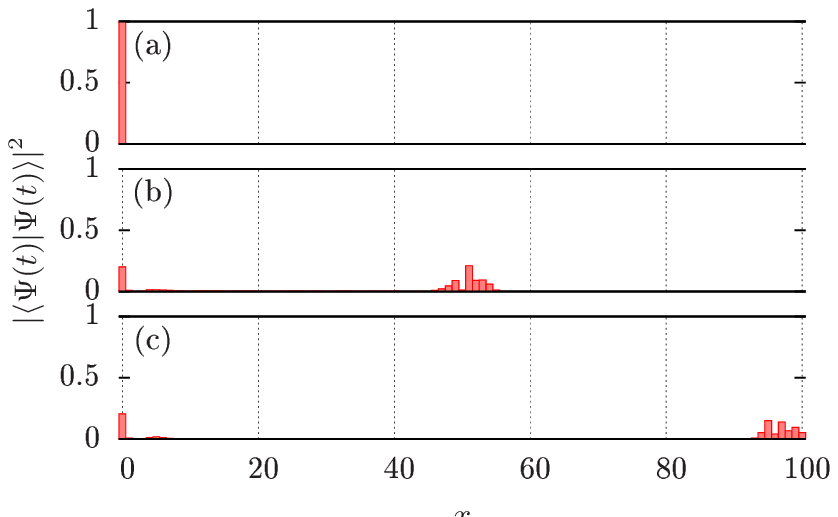}
  \end{center}
  \caption{(Color online) Excitation probability plotted against $x$ for
    a spin chain with $N = 101$ sites, at times (a) 0, (b) 100 and (c)
    200, in units of $J^{-1}$. Here, $d(t) = 0.5 t$ and $C(t) =
    0.1$. The final fidelity here is around 5\%.}
  \label{fig:nocontpop2}
\end{figure}

\begin{figure}[tp]
  \begin{center}
    \includegraphics{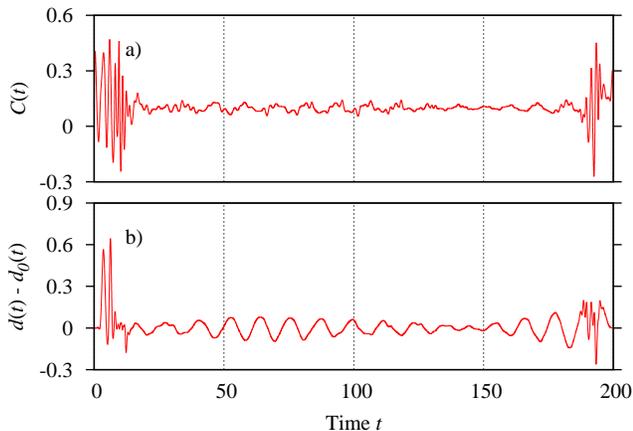}
  \end{center}
  \caption{(Color online) The optimal control pulses for (a) $C(t)$ and
    (b) $d(t) - d_0(t)$, where $d_0(t) = 0.5 t$. The main features here
    are large perturbations at the initial and final time due to the
    boundaries, and slower modulations for the intermediate stage of the
    transport.}
  \label{fig:contpulse}
\end{figure}

Here, we implement an optimal control algorithm most commonly known as
the Krotov method. In outline, the method works as follows.
\begin{enumerate}
\item We solve the Schr\"{o}dinger equation from \eqref{eq:schrod} to
  find \ket{\Psi(T)}, where $T$ is the total evolution time.

\item \label{item:loop} We define the co-state $\ket{\chi(T)} =
  \ket{\Psi(T)}\bracket{\Psi(T)}{\varphi_N}$. This state is propagated
  backwards to the initial time.

\item The initial state is then propagated forward again through time,
  but at each time step we calculate the matrix elements
  \begin{equation}
    \bra{\chi(t)}\frac{\partial H(u_n(t); t)}{\partial u_n(t)}\ket{\Psi(t)}
  \end{equation}
  for the two controls $u_1(t) = d(t)$ and $u_2(t) = C(t)$. The matrix
  elements are then used to update the control functions, which are then
  used to propagate $\Psi(t)$ to the next time step.

\item We can then calculate the fidelity of the transport
  \begin{equation}
    F \equiv |\bracket{\Psi(T)}{\varphi_N}|^2\:,
  \end{equation}
  which tells us how close we were to achieving our goal. (Note that we
  will often refer not to the fidelity, but to the infidelity $I \equiv
  1 - F$.) If we achieve fidelity $F = 1$ (up to a given threshold), we
  stop the optimisation, otherwise we begin again at step
  \ref{item:loop}.
\end{enumerate}
There are several aspects in implementing the algorithm which are
described in more detail in
Ref.~\cite{Sklarz2002}. Figure~\ref{fig:nocontpop} shows the
non-adiabatic transfer of a spin excitation across a chain of $N = 101$
spins without applying optimal control. One sees that during
propagation, much of the spin excitation has been left behind. One way
to correct this would be to lower the field strength: this will allow
neighbouring sites to interact for longer, so that more of the
excitation can be transmitted. However, this causes the excitation to
spread out, which can be seen in Fig.~\ref{fig:nocontpop2}. In
comparison with Fig.~\ref{fig:nocontpop}, we see that although we have
not left as much of the excitation behind, we have spread it over more
sites.%
\begin{figure}[tp]
  \begin{center}
    \includegraphics{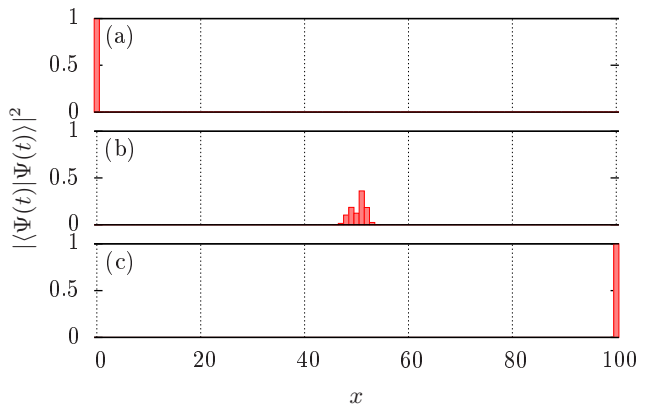}
  \end{center}
  \caption{(Color online) Excitation probability plotted against $x$ for
    a spin chain with $N = 101$ sites, at times (a) 0, (b) 100 and (c)
    200, in units of $J^{-1}$. Here, $d(t)$ and $C(t)$ were optimally
    controlled. The final fidelity is $> 99\%$.}
  \label{fig:contpop}
\end{figure}%
After applying optimal control (300 iterations of the update procedure),
we arrive at the evolution shown in Fig.~\ref{fig:contpop}. Here we see
that we no longer leave excitation behind at the initial spin sites, and
although we spread out the excitation during transport, we successfully
recover the highly localised final state, giving a final fidelity $F$
that differs from unity by $<10^{-4}$. The pulses required to achieve
this result are shown in Fig.~\ref{fig:contpulse}. 
Typical features of these pulses are large modulations at the
boundaries, necessary for `accelerating' (it will be useful here to
imagine an excitation wave) the excitation at the initial time, and then
`decelerating' it near the final time. Small modulations are required at
intermediate times in order to prevent the excitation from spreading
over too many sites. It is also worth noting that the speed achieved
here is at least two orders of magnitude faster than is possible in the
adiabatic case for comparable fidelities \cite{Balachandran2008}.

\begin{figure}[tbp]
  \begin{center}
    \includegraphics{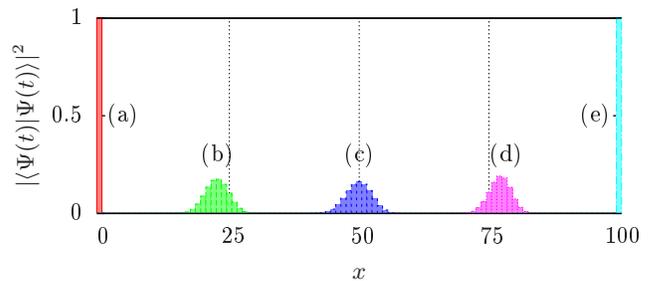}
  \end{center}
  \caption{(Color online) The probability density of the wavefunction
    along the chain at different times: (a) $t = 0$, (b) $t = T/4$, (c)
    $t = T/2$, (d) $t = 3T/4$, and (e) $t = T$, where $T = \tqn =
    56.50J^{-1}$. Both $d(t)$ and $C(t)$ were found after 100,000
    iterations of the optimal control algorithm.}
  \label{fig:oct_at_sl}
\end{figure}

If we decrease $T$, we find for all times $T$ shorter than a particular
time $\tqn$ that even after applying the optimisation algorithm we are
still unable to achieve high-fidelity state transfer. In other words,
there is a minimum time required to perform the transfer
\cite{Caneva2009}. The lower-bound on the value of $\tqn$ is set by the
quantum speed limit (QSL); no transfer can take place faster than the
QSL allows. Fig.~\ref{fig:oct_at_sl} shows the same transfer of
excitation as in Fig.~\ref{fig:contpop}, but in this case we have set
the total allowed time $T = \tqn$ (how we determined $\tqn$ is shown
later). One sees clearly that the evolution of the system is that of a
wave of excitation, moving with an almost constant velocity along the
chain.
\begin{figure}[tbp]
  \begin{center}
    \includegraphics{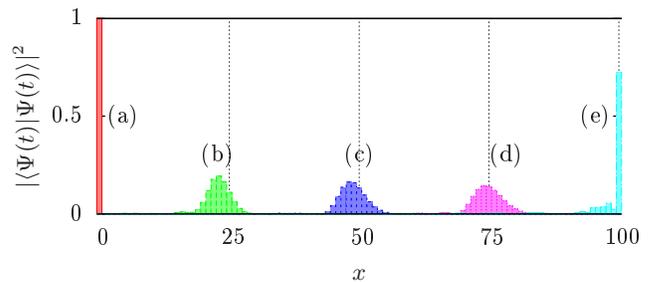}
  \end{center}
  \caption{(Color online) The probability density of the wavefunction
    along the chain at different times: (a) $t = 0$, (b) $t = T/4$, (c)
    $t = T/2$, (d) $t = 3T/4$, and (e) $t = T$, where $T < \tqn =
    53.30J^{-1}$. Both $d(t)$ and $C(t)$ were found after 100,000
    iterations of the optimal control algorithm.}
  \label{fig:oct_past_sl}
\end{figure}
When we choose time $T < \tqn$, we find accordingly that the optimal
control algorithm is unable find an optimal solution, even after many
thousands of iterations. This is a strong indication that we have gone
beyond the quantum speed limit, and there is no solution by which we can
transfer the excitation across the chain in the given time. The
evolution of the system in this case is shown in
Fig~\ref{fig:oct_past_sl}. In comparison to Fig.~\ref{fig:oct_at_sl}, one
sees that the evolution looks much the same. However, if one compares
the excitation profile at $T/2$ for both evolutions, one sees that while
the evolution at the QSL has the excitation wave centred at the 51st
site (i.e.\ the halfway point), the evolution for a time $T < \tqn$
falls short of the halfway point after $T/2$. This is an indication that
we are indeed beyond the QSL, since if we cannot reach the halfway point
before half the time has elapsed, we might well guess that we cannot
reach the final site in the remaining half of the time.

\begin{figure}[tbp]
  \begin{center}
    \includegraphics{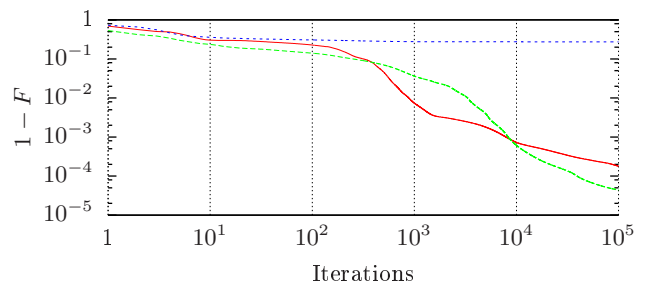}
  \end{center}
  \caption{(Color online) The decrease in infidelity of a transfer
    across a chain with 101 sites against the iterations of the control
    algorithm. The solid (red) line is the convergence for a transfer
    time $T = 70.92J^{-1} > \tqn$, the dashed (green) line for a
    transfer time $T = \tqn = 56.50J^{-1}$, and the dotted (blue) line
    for a transfer time of $T < \tqn = 53.33J^{-1}$.}
  \label{fig:oct_error}
\end{figure}

We can see this failure of the optimisation algorithm more clearly in
Fig.~\ref{fig:oct_error}. %
For times $T > \tqn$, the infidelity converges almost exponentially
towards zero. For times $T < \tqn$, the decrease in infidelity saturates
after several hundred iterations.

Another indication that the QSL has been reached can be found by
examining the average ``velocity'' of the excitation wave as it moves
across the chain. Given a total time $T$ for the propagation, the
average rate at which the excitation \emph{should} be transmitted is
given by $v_a = (N - 1)/T$. Examining the dynamics, we can see that for
much of the propagation time, the excitation moves along the chain with
an (approximately) constant velocity. We can quantify this velocity as
\begin{equation}\label{eqn:actual_speed}
  v_d = \frac{4}{T^2} \int_{\frac{T}{4}}^{\frac{3T}{4}}
  \langle x \rangle \ud t\:,
\end{equation}
where $\langle x \rangle = \bra{\Psi(t)} x \ket{\Psi(t)}$ is the
expectation value of the position of the excitation along the chain. In
other words, we take the average position of the excitation in the time
interval $[T/4,3T/4]$ (to avoid effects at the ends of the chain) and
divide by the average time taken to reach that position, $T/2$.

\begin{figure}[tbp]
  \begin{center}
    \includegraphics{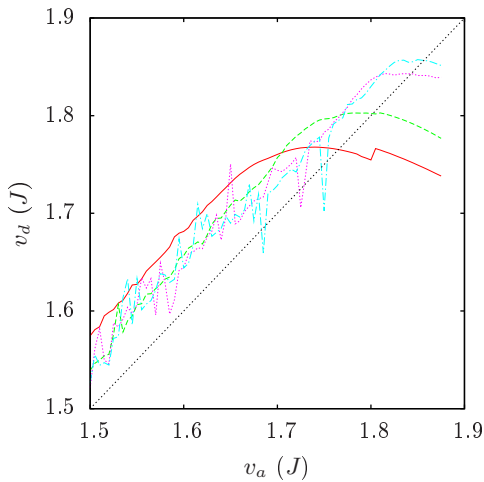}
  \end{center}
  \caption{(Color online) The average speed of the excitation wave $v_d$
    versus $v_a$. The solid (red) line shows the effect of filtering on
    the optimised pulses for a chain length of 41 sites, the long-dashed
    (green) line shows the same for 61 sites, the short-dashed (dark
    blue) line for 81 sites, the dotted (pink) line for 101 sites, and
    the dashed-dotted (light blue) line for 121 sites. The black dotted
    line is the line $v_a = v_d$.}
  \label{fig:oct_wavespeed}
\end{figure}

In the ideal case, we would have $v_a = v_d$, in which case the optimal
solution would be the transit of the excitation along the chain at
exactly the average rate required to reach the other end. However, as we
cross the threshold set by the QSL, we should find that $v_d$ reaches a
maximum, which is the maximum speed at which the excitation can
propagate. This is exactly what is seen in Fig.~\ref{fig:oct_wavespeed}.

\begin{figure}[tbp]
  \begin{center}
    \includegraphics{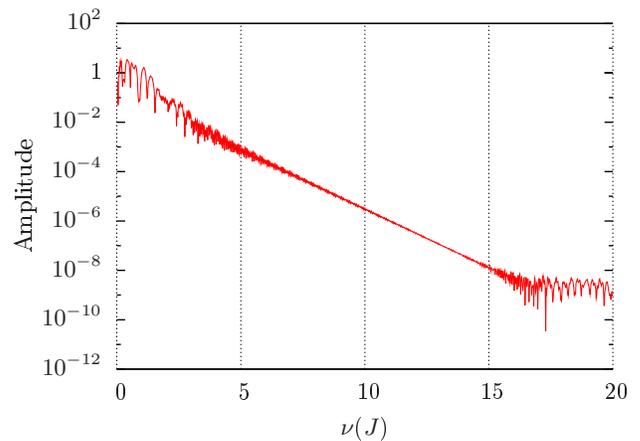}
  \end{center}
  \caption{(Color online) The solid (red) line shows the Fourier
    transform of the pulse $d(t) - d_0(t)$ for a chain length of 101
    spins with a total time $T = 56.50J^{-1}$.}
  \label{fig:ftrans}
\end{figure}

\begin{figure}[tbp]
  \begin{center}
    \includegraphics{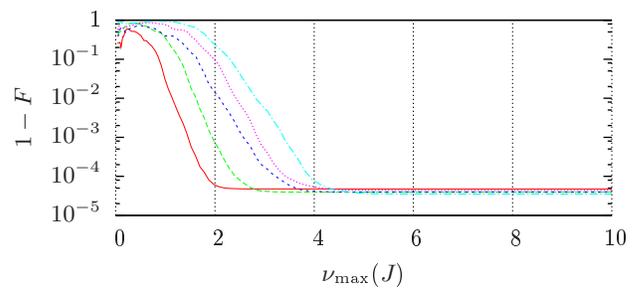}
  \end{center}
  \caption{(Color online) The infidelity of the transfer related to the
    maximum frequency component of the controls retained after
    filtering. For key, see Fig.~\ref{fig:oct_wavespeed}.}
  \label{fig:oct_fourier}
\end{figure}

\begin{figure}[tbp]
  \begin{center}
    \includegraphics{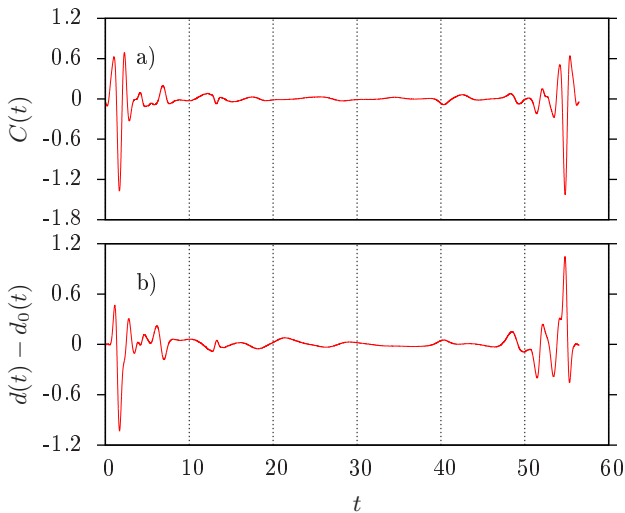}
  \end{center}
  \caption{(Color online) The optimal control pulses for (a) $C(t)$ and
    (b) $d(t) - d_0(t)$, where $d_0(t) = 1.77 t$. The maximum frequency
    component is $\sim4 J$.}
  \label{fig:smoothed_pulses}
\end{figure}

The last issue we want to address is robustness. In essence, how much
information in the control pulses given in Fig.~\ref{fig:contpulse} (and
indeed in all of the control pulses at the QSL) can be discarded without
detriment to the transfer fidelity? Figure~\ref{fig:ftrans} %
shows an example spectrum of a pulse for $d(t)$ for a transfer along a
chain of 101 spins at the QSL, and Fig.~\ref{fig:oct_fourier}
shows the effect on the fidelity after filtering the optimised
pulses. The filter applied is a simple frequency cutoff: the pulse (in
frequency space) is convoluted with a function
\begin{equation}
  \label{eq:conv_func}
  \gamma(\nu; \nu_{\text{max}}) = \begin{cases}
    \nu& \text{if}\ |\nu| \leq \nu_{\text{max}}, \\
    0&\text{otherwise},
  \end{cases}
\end{equation}
where $\nu_{\text{max}}$ is the maximum allowed frequency in the
pulse. We see that not all of the frequencies in the control pulses need
be retained; on average, we only need frequencies up to around $4 J$ in
order to maintain a high fidelity. Note that this is independent of the
chain length $N$. Figure~\ref{fig:smoothed_pulses} shows a set of pulses
that transport the excitation along a chain of 101 spins with an
infidelity $I<10^{-4}$, where the maximum frequency component is
$\sim 4J$.


\section{Theoretical limits of non-relativistic quantum
  theory}\label{sec:conn}

Does the limit $\tqn$ discussed in the last chapter have a physical
origin, or is it simply a numerical constraint, stemming from the
construction of the optimisation routine itself? If, in fact, we
\emph{are} able to reach the physical limit by application of optimal
control routines, then it would appear that optimal control can not only
be used to improve the operation of experimental implementations, but
indeed to probe a system's dynamics and physical limits. This connection
was already investigated in Ref.~\cite{Caneva2009}; here, we elucidate
further the methods that were applied and more specific conclusions.

The physical limits on quantum systems (and hence any physical system)
have been investigated theoretically in quantum systems for several
years; such considerations lead Lloyd in 2000 to calculate the maximum
rate at which any machine can process information \cite{Lloyd2000}. In
particular, the notion of a ``quantum speed limit'' has been reported by
several authors. We briefly recount this theory and its particular
application to our problem.

\subsection{The quantum speed limit}\label{sec:qsl}

What is the absolute maximum speed at which we can transfer information
along our chain? This amounts to finding the minimum time it takes for
the given initial state $\ket{\Psi(0)}$ to evolve to the goal state
$\ket{\varphi_N}$. A possible route for finding this minimum time was
explored by Carlini \emph{et al.} \cite{Carlini2006}, where it was shown
that one may derive the time-optimal Hamiltonian for a given state
evolution by minimising the quantum action $S$ of the system, by which
the problem may be interpreted as a quantum analogue of the classical
brachistochrone. In principle, the same procedure could be performed in
our case, but the complexity of the calculation is prohibitive for a
many-body system like ours. Hence we ask a somewhat simpler question, as
in Giovannetti \emph{et al.} \cite{Giovannetti2003}: how fast can a
quantum system under a time-independent Hamiltonian evolve in time?

The notions of energy and time are not inseparable, an idea that
presents itself in the enigmatic time-energy uncertainty relation
\cite{Busch2008}. Hence the minimum time in which we can perform some
given evolution must be connected to the related energy scales. This
minimum time is referred to as the quantum speed limit (QSL). For the
case where the evolution is from an initial state to an orthogonal state
for a time-independent Hamiltonian, this relation can be written
explicitly as \cite{Giovannetti2004}
\begin{equation}
  \label{eq:tqsl_static}
  \tqa \equiv \max\left(\frac{\pi \hbar}{2
      E},\frac{\pi \hbar}{2 \Delta E}\right)\:,
\end{equation}
with
\begin{gather}\label{eq:del_e}
  \Delta E \equiv \sqrt{\bra{\psi(0)}[\hat{H}(t) - E
    (t)]^2\ket{\psi(0)}}\:,\\
  \label{eq:ela}
  E \equiv \bra{\psi(0)}\hat{H}(t)\ket{\psi(0)}\:.
\end{gather}
As pointed out, this is only valid when the time evolution is governed
by a time-independent Hamiltonian: $E$ and $\Delta E$ are a measure of
the energy resources available in the system only at the initial time,
which for time-independent Hamiltonians defines a fixed energy scale. In
our case, the methodology must be slightly modified, by considering
instead the \emph{mean} energy spread of our system as it evolves under
our time-dependent Hamiltonian, which we find by averaging the
instantaneous energy spread of the system over the time interval
$[0,T]$. By integrating over time, we effectively apply the bound to
infinitesimal time steps $\ud t$ where the Hamiltonian is approximately
constant. We modify the definition in Eq.~\eqref{eq:tqsl_static} to read
\cite{Caneva2009}
\begin{equation}
  \label{eq:tqsl_av2}
  \tqa \equiv \max\left\{\frac{\pi\hbar}{2J}, \frac{\pi \hbar}{2 \Delta\mathcal{E}_2}\right\}\:,
\end{equation}
where
\begin{equation}
  \Delta\mathcal{E} = \frac{1}{T}\int_0^T \Delta E (t) \ud
  t\:,
\end{equation}
with
\begin{gather}\label{eq:del_elam}
  \Delta E (t) \equiv \sqrt{\bra{\Psi(t)}[\hat{H}(t) - E
    (t)]^2\ket{\Psi(t)}}\:,\\
  \label{eq:elam}
  E (t) \equiv \bra{\Psi(t)}\hat{H}(t)\ket{\Psi(t)}\:.
\end{gather}
As was already pointed out, this speed limit defines the time it takes
to rotate from the initial state to an orthogonal state. Since the
initial and final sites are not directly coupled, we cannot immediately
rotate from the initial state to our goal state. Due to this condition,
we postulate that the speed limit must be interpreted as an effective
\emph{time-per-site}; the total time it takes to traverse the chain is
this time-per-site multiplied by the number of sites (minus one) in the
chain, or equivalently, the number of edges we have between the initial
state vertex and the final state vertex when one views the spin chain as
a connected graph.

Equation \eqref{eq:tqsl_av2} effectively states that the minimum time it
takes to rotate from the current system state to an orthogonal state is
bounded from below by $\pi \hbar / (2 J)$ (we shall see later that for
the evolutions we consider, the second term in Eq.~\eqref{eq:tqsl_av2}
is always less than this term, so that we can neglect it). By
considering the speed limit of a simple two-spin system with a coupling
strength $J$, we can associate this bound with the time it takes to swap
an excitation between only two sites, given that for the initial state
the excitation is completely localised on one of the two sites. Using
the reasoning above, we see that the quantum speed limit theory predicts
that the minimum time to traverse the chain is given simply by the time
it takes to perform a swap between two neighbouring sites (which we
shall henceforth refer to as `orthogonal swaps') mutiplied by the number
of sites in the chain (minus one). However, in our particular system, at
some intermediate time it may be (as we have already seen from the
results in Section~\ref{sec:oct}) that the excitation does not perform
repeated swap operations, but rather moves along the chain as a
dispersed ``wave''. If one now imagines the picture of the excitation
wave moving from site to site, we note that two excitation waves centred
at neighbouring sites are not orthogonal, unlike when we have the
excitation fully localised on a single site. This means that we can
expect the actual propagation time to be \emph{shorter} than the one
calculated from simply doing repeated orthogonal swap operations. The
optimised system performs a controlled excitation-wave propagation,
which we can view as a cascade of \emph{effective} swap operations, each
shorter in duration than that given by the orthogonal swap. We are then
motivated to write the total time to traverse the chain as
\begin{equation}
  \label{eq:fullqsl}
  T_{\mathrm{QSL}} = \gamma (N - 1) \tqa\:,
\end{equation}
where $\gamma$ is a dimensionless constant that quantifies the effective
swap duration in terms of the orthogonal swap. As a side remark, we note
that one can also imagine mapping the full chain with the effective
swaps onto a shorter chain with orthogonal swaps, which is analogous to
a reduction of the transmission length of the chain. Similar ideas have
already been explored for long range interactions in
Ref.~\cite{Gualdi2008}.

\subsection{Comparing limits}\label{sec:comp}

As already alluded to in Section~\ref{sec:oct} there comes a point where
the optimal control algorithm is no longer able to reach an optimal
solution. We aim to show that this limit on the evolution time (which we
denoted by $\tqn$) corresponds to the quantum speed limit for the system
$\tqac$ discussed in Section~\ref{sec:qsl}.

\begin{figure}[tbp]
  \centering
  \includegraphics{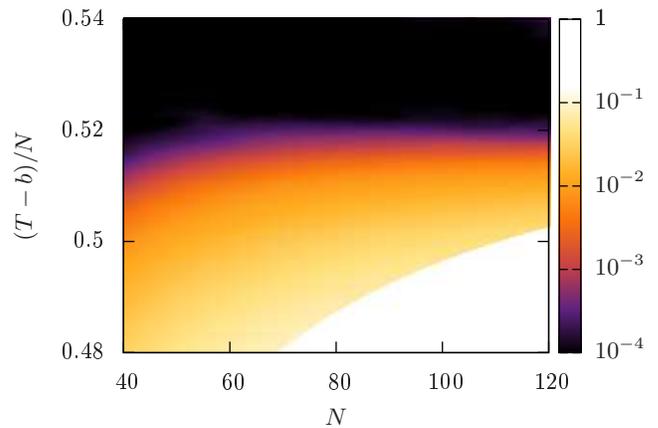}
  \caption{(Colour online.) The infidelity reached after $R = 100,000$
    iterations for different chain lengths $N$ and effective
    time-per-site $(T - b)/(N - 1)$, where $b = 3.65$.}
  \label{fig:qsl_data}
\end{figure}

The procedure for determining $\tqn$ is as follows. We select a chain
length $N$, and set some initial evolution time $T$ which we assume to
be longer than the corresponding $\tqn$. We perform optimal control on
the system for a fixed number of iterations $R$. We then repeat this for
shorter and shorter times $T$. The results of the simulations are shown
in Fig.~\ref{fig:qsl_data}. Note that we plot the effective
time-per-site $(T - b)/(N - 1)$ in order to make comparisons between
chains of differing lengths easier. One sees clearly that for longer
times, we are able to complete the state transfers with high
fidelities. As we reduce the time, we begin to see that the final value
of the infidelity does not converge to zero, even after many thousands
of iterations of the control algorithm. Somewhere in between these two
extremes lies the limit of the optimal control algorithm. We quantify
this by setting a threshold $\varepsilon$ for the infidelity; the time
$\tqn$ for each $N$ is defined as the smallest value of the time $T$ for
which the infidelity $I < \varepsilon$ after $R$ iterations. This
threshold obeys a linear relation:
\begin{equation*}
  \tqn \approx a (N-1) + b
\end{equation*}
with $a = 0.34$ and $b = 3.65$. Note that this is \emph{a posteriori}
the same $b$ used in the effective time-per-site $(T - b)/(N - 1)$ for
Fig.~\ref{fig:qsl_data}. The introduction of the constant $b$ describes
additional effects due to the boundaries of the chain, where the
excitation wave is generated at the beginning of the evolution, and then
collapsed into a localised excitation at the end. Additionally, $b$ is
not dependent on $N$ (unless the chain lenght is of the order of the
width of the spin-wave).

\begin{figure}[tbp]
  \centering
  \includegraphics{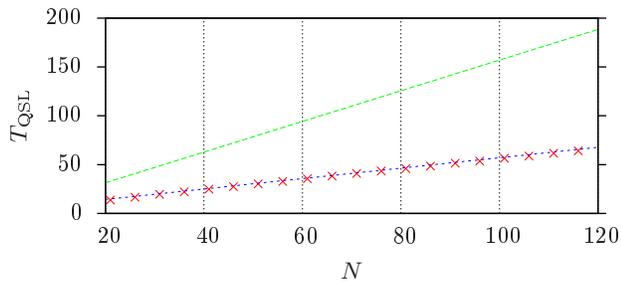}
  \caption{(Colour online.) A comparison of the quantum-speed-limit-time
    $\tqac$ with the optimal control limit $\tqn$. The solid (red) line
    is $\tqac$ with $\gamma = 1$, which is the repeated orthogonal
    swaps. The (blue) crosses are $\tqn$ for different $N$ in the range
    21--131 with $\varepsilon = 5\cdot 10^{-5}$. The dashed (green) line
    is $\tqac + b$ with $\gamma = 0.34$.}
  \label{fig:qsl_oct_comp}
\end{figure}

We now compare the results from the quantum speed limit $\tqac$ with
$\tqn$, which is shown in Fig.~\ref{fig:qsl_oct_comp}. In order to
evaluate Eq.~\eqref{eq:tqsl_av2} for each value of $N$, we must
numerically calculate the second term in the bracket, since it depends
upon the time evolved state $\ket{\psi(t)}$. For all points at our
defined threshold, this comes out to be less than the first term in the
bracket in Eq.~\eqref{eq:tqsl_av2}, so that the speed limit is given
simply by the effective swap time. One finds that optimal control
outperforms what can be achieved through applying repeated swap
operations between adjacent spins. Furthermore, by ignoring boundary
effects for $\tqn$, we find that our model for the quantum speed limit
fits the data with a value of $\gamma = 0.34$. This means that the speed
limit achieved with the optimal control can be described (ignoring the
ends of the chain) as a cascade of effective swaps.


\section{Conclusion}\label{sec:conc}

We have shown that we can successfully apply optimal control to the
system given in Eq.~\eqref{eq:ham} to produce fast transfers of
excitations along spin chains; two orders of magnitude faster, in fact,
than was reported in Ref.~\cite{Balachandran2008} for comparable
fidelities. This has application in the fast transport of quantum states
over short distances. Furthermore, we have found a fundamental limit for
optimal control beyond which optimisation is not possible, and identified
it as a speed limit on the dynamics of the system, which is manifested
by the dynamics as the propagation of an excitation wave with constant
velocity. We compare this with the standard formulation of the quantum
speed limit, and show that for our many-body problem, the quantum speed
limit implies that the optimal strategy for transport is characterised
by effective swaps along the chain. We confirm this through a comparison
with the numerical results.

It is interesting to note that aside from the theory on the quantum
speed limit, there is a large body of work concerned with a similar
bound specifically for spin systems, namely the Lieb-Robinson bound
\cite{Lieb1972,Robinson1976,Bravyi2006,Hamma2009}. It would be
interesting to investigate the connection between this bound and the
QSL, although it is likely difficult to quantify this explicitly.

We have shown that not only is optimal control a useful tool for the
optimisation of tasks relevant for quantum information processing
(specifically transmission of quantum information along a spin chain),
but also as a means to probe the limits of many-body quantum systems
where the theoretical methods become unwieldy. We expect that given the
generality of the method, it should be able to probe fundamental limits
of many quantum systems that can be efficiently simulated. Indeed, we
used the same technique to prove a bound on the duration of a unitary
\textsc{swap} operation on a spin chain, showing that it was achievable
in a time that scaled only polynomially with the number of sites
\cite{Burgarth2009} (although it was not shown that this was a
fundamental limit). We will continue with such investigations in future
work.

We would like to thank L. Viola for valuable discussions. We acknowledge
financial support by the EU under the contracts MRTN-CT-2006-035369
(EMALI), IP-EUROSQIP, IP-SCALA, an IP-AQUTE, and from the German SFB
TRR21. We thank the bwGRID for computational resources.


\bibliography{reduced}

\end{document}